\documentclass[preprint,12pt]{elsarticle}

\usepackage{amsmath,amssymb,amsfonts}
\usepackage{epsfig}
\def \o{\omega}

\def \be{\begin{equation}}
\def \ee{\end{equation}}

\def \r{\mathbf{r}}
\def \k{\mathbf{k}}
\def \kk{\boldsymbol \kappa}

\def \I {\hbox{\rm Im}}

\journal{superlattices and microstructures}
\begin{document}
\begin{frontmatter}

\title{Weak and strong coupling of a quantum emitter with a meta--surface}

\author{Didier Felbacq}
\address{
Universit\'e de Montpellier 2, Laboratoire Charles Coulomb\\ Unit\'e Mixte de Recherche
du Centre National de la Recherche Scientifique 5221\\ 34095 Montpellier, France}

%\author{Didier Felbacq}
%\affiliation{Universit\'{e} de Montpellier II\\Laboratoire Charles Coulomb \\
%Unit\'e Mixte de Recherche du Centre National de la Recherche Scientifique 5221\\
%34095 Montpellier Cedex 5, France\\}
%\authorinfo{Further author information: (Send correspondence to D. Felbacq)\\D. Felbacq: E-mail: Didier.Felbacq@univ-montp2.fr, Telephone: +33 (0)467 143 216}

\begin{abstract}
Meta--surfaces are the bidimensional analogue of metamaterials. They are made on resonant elements periodically disposed on a surface. They have the ability of controlling the polarization of light and to generalized refraction laws as well. They have also been used to enhance the generation of the second harmonic.
It seems however that their near-field properties have not been investigated. In this work, the coupling of an emitter with a meta--surface made of a periodic set of resonant linear dipoles was studied. Bloch surface modes localized on the meta--surface exist due the resonance of the dipoles. The strong coupling regime with a emitter can be reached when the Bohr frequency of the emitter is in resonance with the Bloch modes of the meta-surface.
\end{abstract}

\end{frontmatter}

\section{Introduction}
%The various coupling regimes are exhibited and a quantum formalism is provided.
Meta--surfaces are the 2D analogue of metamaterials  \cite{metsh}. They are made of basic, resonant, elements disposed on a surface. For electromagnetic waves whose wavelength is larger than the period, the basic elements behave collectively and provide new means of controlling the flow of light  \cite{capasso}. Meta-surfaces are generally seen as devices able to control the far-field behavior of light, such as the polarization state, the directivity, the light-by-light manipulation or the generation of second harmonic signal  \cite{alunature}. Some have made claims that they made possible "generalized laws of diffraction" as compared to Snell-Descartes laws  \cite{capasso}. However, because of their resonant properties, meta--surfaces also have interesting properties in the near-field.
In the present work, we aim at initiating the study of the quantum electrodynamics of meta-surfaces  \cite{quantmat}. In standard cavity quantum electrodynamics, one studies the coupling between an emitter, such as an atom, or quantum dot or a superconducting qubit  \cite{alexZ}, and the electromagnetic modes. Depending on the ratio between the light-matter coupling and that to the irreversible mechanisms, two regimes can occur: the weak coupling and the strong coupling. In the weak coupling regime the losses dominate and the spontaneous decay rate of the emitter is modified by the structured electromagnetic field. This is essentially the Purcell effect. In the strong coupling regime, the coupling dominates the losses: the quantum emitter and the meta-surface form a quantum system whose behavior cannot be decoupled between two separated objects. Rather, the emitter and the meta-surface can exchange photons periodically in time, which leads to hybrid excited states.  From an experimental point of view, this regime leads to the onset of a double peak in the emitted spectrum, due to the anti-crossing of the dispersion curves of the light and matter modes. This situation has been observed in cavity with hybrid states between photons and excitons  \cite{weisbuch} as well as between photons and plasmons  \cite{bellessa}. It was recently predicted theoretically that the strong coupling could be reached between a quantum emitter and Anderson localized modes  \cite{cazcarm}.
In the present work, the coupling of a quantum emitter with the photonic surface modes supported by a meta--surface is investigated. The meta--surface is made of a periodic set of parallel nano wires. From a theoretical point of view, the meta-surface can be described by an effective impedance model, which allows to derive the density of electromagnetic modes due to the meta-surface. Further, it allows to obtain the dressed susceptibility of the quantum emitter and to exhibit the strong coupling regime. An {\it ab initio} numerical simulation of the meta-surface (containing a finite number of nano wires) is used in order to simulate the various regimes.
\begin{figure}
   \begin{center}
   \includegraphics[height=8cm]{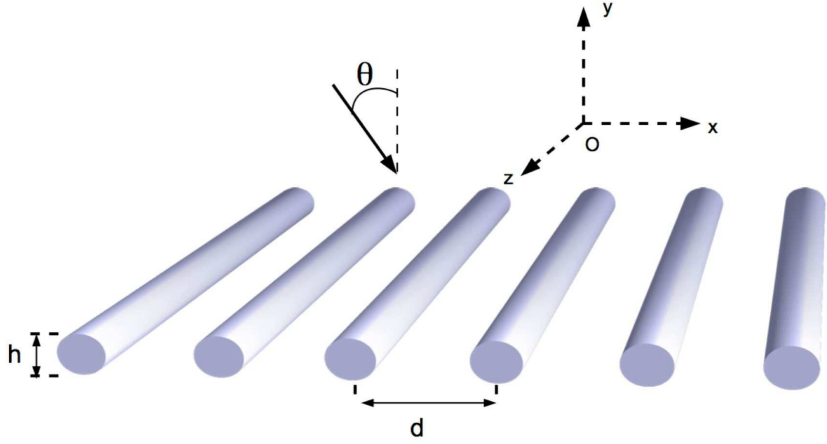}
   \end{center}
   \caption
   { \label{fig1} Sketch of the structure under study.}
\end{figure}
\section{Impedance operator description of the meta--surface}
The nano wires are disposed periodically with a period $d$.
They show a resonant behavior at frequency $\omega_0$: they are described by a dipolar susceptibility that has a non-zero component $S_0(\omega)$  \cite{josa} along the axis of the wires only. Therefore, the only relevant polarization is $E_{||}$, that is, with the electric field parallel to the axis of the nano wires (Possible experimental realizations are discussed in section \ref{numer}). When the collection of nano wires is illuminated by a plane wave $e^{i(kx-Ky)}$, it gives rise to a scattered field that can be written:
\begin{equation}
U^s(\r)=\sum_{m}  s^0 e^{ikmd} \varphi_0(\r-md e_x).
\end{equation}
where $\varphi_0(\r)=H^{(1)}_0(k_0|\r|)$, and $H^{(1)}_0$ is the 0$^{th}$ Hankel function of order $1$. From multiple scattering theory  \cite{PNFA,colrec}, it can be shown that:
\begin{equation}\label{s0}
s^0(\o,k)=\left[1-S_0(\o)\Sigma_0(\o,k)\right]^{-1}S_0(\o),
\end{equation}
where  the lattice sum  \cite{lipton} $\Sigma_0$ is given by  \cite{F3,bloch}:
\begin{eqnarray}\label{sig0}
\Sigma_0(\o,k)=\sum_{m \neq 0}  e^{ikmd}\varphi_0(md).
%=-1-\frac{2i}{\pi}\gamma+\frac{2i}{\pi}\ln\left(\frac{4\pi}{k_0 d}\right)+\frac{2}{d\beta_0(\omega,k)}\\
%+\frac{2}{d}\sum_{n>0} \left( \frac{1}{\beta_n(\omega,k)}+\frac{1}{\beta_{-n}(\omega,k)}-\frac{d}{i\pi |n|}\right).
\end{eqnarray}
The diffracted field can be put in the form of a Rayleigh series, familiar from grating theory \cite{roger}:
\begin{equation}\label{scant}
U^s(x,y;\omega,k)=
\frac{2 s^0_0(\omega,k)}{d} \sum_n \frac{1}{K_n(\omega,k)} e^{i(k_n x+K_n(\omega,k) |y|)}.
\end{equation}
where $k_n=k+\frac{2n\pi}{d}$ and $K_n(\omega,k)=\sqrt{(\omega/c)^2-k_n^2}$.
This expression shows that the field diffracted by the meta-surface is a finite sum of propagative plane waves, and an infinite sum of dissociated plane waves, that is, evanescent away from the meta-surface. The directions of the propagative plane waves are given by the usual relation for diffracted orders: $\sin \theta_n-\sin \theta=n \frac{\lambda}{d}$, irrespectively of the existence of resonances inside the nano-wires. This shows that there is no such thing as a ``generalized law of diffraction'' that would hold for meta-surfaces, as claimed by some  \cite{capasso}.

Formally, this field is the response of the meta-surface to an incident plane wave. It is the symbol of the pseudo-differential operator that represents the link between an arbitrary incident field and the diffracted field. For an incoming field that is a superposition of plane waves in the form: $U^i(\r)=\int \hat{U}(k) e^{i \k \cdot \r}dk,\, \k=(k,-K)$ and $k^2+K^2=k_0^2$, the field diffracted by the meta-surface is given by:
\be
E^g(\r)=\int dk \, U^s(\r;\omega,k) \, \hat{U}(k)
\ee

The boundary conditions at $y=0$ are  \cite{PNFA, colrec}:
$$
[U(x,0^+)-U(x,0^-)]=0,\,  \left[\frac{\partial U}{\partial y}(x,0^+)-\frac{\partial U}{\partial y}(x,0^-)\right]=V(x)\,,
$$
where:
$
V(x)=2i\sum_n \beta_n U^{s}_n e^{ik_n x}
$
%and $r(\alpha)=\frac{2 b_0}{d \beta_0}\sum_n \delta(\alpha-\alpha_n)$.
These conditions can be rewritten conveniently in the operator form:
\begin{equation}
Z_0
F^+=F^-  \,,
\end{equation}
where:
$F^+=\left( \begin{array}{c} U(x,0^+)  \\ \frac{\partial U}{\partial y}(x,0^+)\end{array}\right)$, $F^-=\left( \begin{array}{c} U(x,0^-)  \\ \frac{\partial U}{\partial y}(x,0^-)\end{array}\right)$ and $Z_0$ is the transfer matrix of the meta surface.
When the wavelength is large enough, the evanescent fields can be neglected and the meta-surface can be described by a simple $2 \times 2$ matrix:
$$
Z_0=\left(
\begin{array}{cc}
 1  & 0 \\
 \frac{-2ik_0 r}{r+1} &   1
\end{array}
\right)
$$,
where $r=\frac{2 s^0_0(\omega,k)}{K\,d}$. It behaves as an infinitely thin current sheet with a conductivity proportional to: $ \frac{-2ik_0 r}{r+1}$. This corresponds to an effective description of the complicated set of nano wires that comes under homogenization theory  \cite{alex,F18b,F13}.
The transfer matrix can be used to describe the electromagnetic behavior of a meta-surface deposited on a substrate \cite{quasi,ultra}.

The Bloch modes that can exist in the structure \cite{F15} are obtained as solutions of Maxwell equations in the absence of an incident field. They correspond to zeroes of $1-S_0\Sigma_0$. The dispersion relation is therefore the set of couples $(\omega,k) \in \mathbb{R}^+ \times ]-\frac{\pi}{d},\frac{\pi}{d}]$ such that: \begin{equation}\label{dips} S_0(\omega)\Sigma_0(\omega,k)=1 \,.\end{equation}

Finally, we determine the Green function $g(\r,\r')$. It is the response of the system to a point source $\delta(\r-\r')$. Let $g_0$ denote the Green function in vacuum: $g_0(\r,\r')=-\frac{i}{4} H_0^{(1)}(k_0|\r-\r'|)$. From Weyl formula  \cite{abramowitz}, the plane wave expansion is obtained: $g_0(\r,\r')=\frac{1}{4i\pi}\int \frac{1}{K} e^{i[k (x-x')+K |y-y']}dk$. The above derivation leads to:
\begin{eqnarray}
g(\r,\r';\o)=g_0(\r,\r')+\frac{1}{4i\pi}\int \frac{1}{K} U^s(\r;\o,k) e^{-i \k \cdot \r'}dk,
\end{eqnarray}

%The expression under the integral is integrable near $\alpha=\pm k_0$, indeed the quantity $\frac{b_0^0(\omega,k)}{\beta_0}$ is bounded in the vicinity of $\omega=kc$.

The cross density of states  \cite{cazcarm} is then given by: $\rho(\r,\r')=-\frac{1}{\pi} \Im G(\r,\r')$.
%$$
%\Im g(\mathbf{r},\mathbf{r}';\omega)=-\frac{1}{4} J_0(k_0 |\mathbf{r}-\mathbf{r}'|)-\int_0^{+\infty} \frac{1}{d} \sum_n \cos(\alpha x) \Im \left(\frac{b_0^0}{i\pi \beta_n}e^{i\beta_n |y|}\right)d\alpha,
%$$

\section{Coupling of the dipole with the meta--surface}
Let us now consider the coupling of a quantum emitter with the meta-surface. The quantum emitter is basically a two-level system, that can be described as a dipole with a susceptibility \cite{cazcarm} $s_D(\o)=\frac{2c^2}{\omega^2}\frac{\Gamma_s^R}{\o_0-\o-i(\Gamma_s^R+\Gamma_s^{NR})/2}$, where $\Gamma_s^R$ and $\Gamma_s^NR$ are, respectively, the radiative and intrinsic non-radiative linewidth. The emitter is assumed to have a resonance at the same frequency $\omega_0$ as the nano-wires.
When the emitter is not too close to the meta-surface, the physical phenomena at stake is the Purcell effect  \cite{chiao}, which is described by means of Fermi golden rule. The spontaneous decay rate is given by:
$$
\Gamma_{ms} \sim \frac{2 \o^2}{\hbar c^2}\,  \I [g(r,r,\o)]
$$
up to an irrelevant factor.
By normalizing by the decay rate in vacuum, one introduces the Purcell factor, defined as:
$F_P=\rho/\rho_0=4 \I [g(r,r,\o)]$. It describes the enhancement of the spontaneous decay rate of a quantum emitter due to its interaction with the structured electromagnetic field produced by the meta-surface.

Let us now study the strong coupling regime. In the celebrated situation that exists in cavity quantum electrodynamics, excitons and photons can form polaritons; here the emitter plays the role of the exciton (flat dispersion curve), while the surface modes of the meta-surface play the role of the electromagnetic cavity modes. The system 'emitter+meta-surface' is illuminated by an incident field $E^i(\r)=\int A^i(k) e^{i \k \cdot \r}$.
The field radiated by the emitter at position $\r_D=(x_D,y_D)$ is given by:
$E^s_D(\r)=b_D(\omega)  H_0(k|\r-\r_D|)$.
Here $b_D(\omega)$ is the scattering amplitude, it is given by: $b_D=s_D \times E_{\rm local}(\r_D)$, where $E_{\rm local}(\r_D)$ is the local field at the position of the emitter: it is the sum of the incident field and that reflected by the meta-surface. Using Weyl formula, $E^s$ can be decomposed into plane waves:
$E^s_D(\r)=b_D(\o) \frac{1}{\pi} \int \frac{1}{K} e^{i \left[k(x-x_D) +K |y-y_D|\right]} dk$.
In order to determine $b_D$, one has to find the expression of the local field. Let $E_g(\r_D)$ denote the field emitted by the meta-surface at the position of the emitter. It is the response of the meta-surface to both the incident field and the field emitted by the emitter.

The incident field impinging on the meta--surface is given by $$b_D(\omega) H_0(k|\r-\r_D|)+E^i(\r)=\int \left[\frac{b_D(\o)}{\pi K}e^{i \kk \cdot \r_D}+A^i(k)\right]e^{i \k \cdot \r} dk.$$ Hence, we obtain:
$E_g(\r)=\int  U^s(\r;\o,k) \left[\frac{b_D(\o)}{\pi K}e^{i \kk \cdot \r_D}+A^i(k)\right]e^{i\kk \cdot \r} dk$, where $\kk=(k,K)$. Consequently, the local field on the emitter is:
 \begin{eqnarray*}
 E^i(\r_D)+ E_g(\r_D)=b_D(\o)\int  dk  \frac{U^s(\r;\o,k)}{\pi K}e^{2i \kk \cdot \r_D}+\\
 \int dk A^i(k)e^{i kx_D} \left[U^s(\r;\o,k)  e^{iK y_D} + e^{-i Ky_D}\right]
 \end{eqnarray*}
 It holds: $b_D(\omega)=s_D(\omega) [E^i(\r_D)+ E_g(\r_D)] $, and therefore
%$$
%b_D=s_D E^i(r_D)+s_D   \int  r(\alpha)(\frac{b_D}{\pi \beta}e^{i\beta y_D}+A(\alpha))e^{i(\alpha x_D+\beta y_D)} d\alpha]
%$$
%$$
%b_D(\omega)=s_D(\omega) b_D(\omega)  \int  r(\alpha) \frac{1}{\pi \beta}e^{i(\alpha x_D +2\beta y_D)} d\alpha+s_D(\omega) \int (1+r(\alpha)) A(\alpha) e^{i(\alpha x_D+\beta y_D)} d\alpha
%$$
we obtain the dressed scattering coefficient:
%$$
%b_D-s_D b_D  \int  r(\alpha) \frac{1}{\pi \beta}e^{i(\alpha x_D +2\beta y_D)} d\alpha=s_D \int (1+r(\alpha)) A(\alpha) e^{i(\alpha x_D+\beta y_D)} d\alpha
%$$
\be
b_D(\omega)=\frac{s_D(\omega) \int dk A^i(k)e^{i kx_D} \left[U^s(\r;\o,k)  e^{iK y_D} + e^{-i Ky_D}\right]}{1-s_D(\omega)\int  dk  \frac{U^s(\r;\o,k)}{\pi K}e^{2i \kk \cdot \r_D }}
\ee

%\subsection{Quantum formalism}
%
%Consider the interaction hamiltonian for the electromagnetic field  and matter in the interaction representation:
%$$H_I=-(\bold{d} \sigma^++\bold{d}^* \sigma^-) \otimes (\bold{E}^++\bold{E}^-)$$
%$$
%H_F=\int_{Y^*}  \sum_p \hbar \omega_p(k) \psi_p(k,r)^{\dag} \psi_p(k,r) dk+\hbar \omega_0 \sigma_z+
%\sigma^+ \otimes E+\sigma^- \otimes E^{\dag}
%$$
%where $\bold{E}^-,\bold{E}^+$ are operator-valued functional. Given a classical wave packet $A(x,t)$,
% $\bold{E}^+[A]$ is operator in the Fock space of finite energy wave packets ${\cal F}=\bigoplus_n \bigotimes^n_{m=1} {\cal H}$, where $ {\cal H}=L^2(\mathbb{R}^3)$. The total quantum phase space is $\mathbb{C}^2 \otimes {\cal F}$.
% The modes of the vacuum and the losses inherent to the meta--surface are considered a reservoir and described quantum mechanically by the Liouvillian ${\cal L[\rho]}$ in the master equation describing the quantum dynamics of the emitter.
% Let us assume that the system is prepared in the state $\left| 1, A \right>$, that is, the quantum emitter is excited and there is an incoming Bloch wave packet on the meta--surface.

\section{Numerical simulations \label{numer}}
\begin{figure}
   \begin{center}
   \includegraphics[height=8cm]{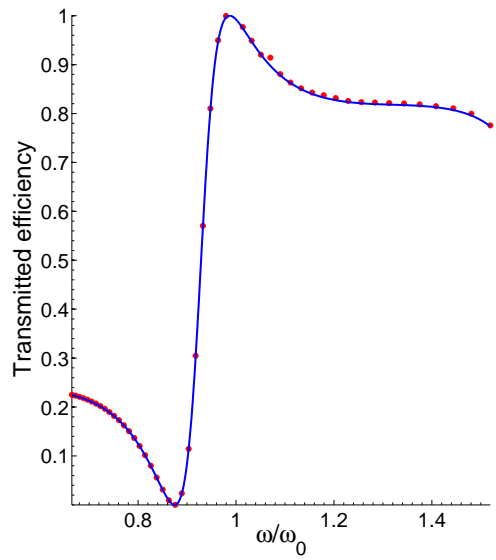}
   \end{center}
   \caption
   { \label{effreso} Energy transmitted in the specular order for an infinite meta-surface (continuous line) and transmission spectrum through the finite meta-surface (stars). The transmission is defined as the normalized flux of the Poynting vector through a segment situated below the meta-surface. The structures are illuminated at normal incidence.}
\end{figure}
In this section, we illustrate the preceding formalism in the case of a 2D resonant emitter situated in the vicinity of a meta surface made of dielectric nano wires. The relative permittivity of the nano wires is chosen to be $12$. The resonant behavior is linked to a Mie resonance that exist at a wavelength $\lambda/a\sim 5.3$, where $a$ is the radius of the nano wires. This can be achieved in the visible domain for nano wires having a diameter of about $200$nm.
Let us start by describing the optical properties of the meta-surface. As a periodic object, it behaves basically as a grating. When illuminated in  normal incidence, the transmitted efficiency (the energy transported in the specular order) as predicted by eq. (\ref{scant}) is given in fig.\ref{effreso} as a continuous line. A Fano--like behavior is seen, with the existence of a zero in the vicinity of the maximum of the efficiency in the first order. These results are obtained for the ideal meta-surface, i.e. for an infinite number of nano wires. From an experimental point of view, only a finite number of nano wires are allowed. Numerically, this can be handled by means of a multiple scattering approach  \cite{josa}. In fig. \ref{effreso}, the transmission spectrum of a meta-surface with $40$ nano wires is given (red markers). An excellent fit with the prediction of the theory is found, showing that the finite meta-surface is very well described by the impedance model.

\begin{figure}
   \begin{center}
   \includegraphics[height=8cm]{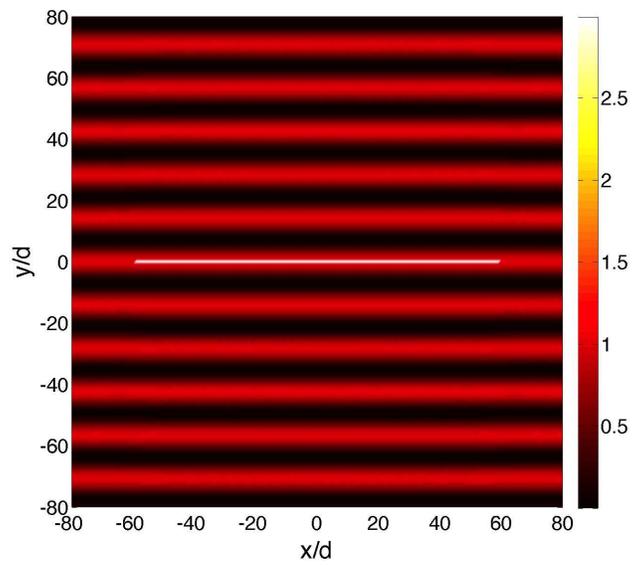}
   \end{center}
   \caption
   { \label{parfaitrans} The finite meta-surface is illuminated in normal incidence, from above, by a plane wave at $\omega/\omega_0=0.99$. The map of the real part of the electric field is plotted. The map is given in false color with a linear scale.}
\end{figure}
\begin{figure}
   \begin{center}
   \includegraphics[height=8cm]{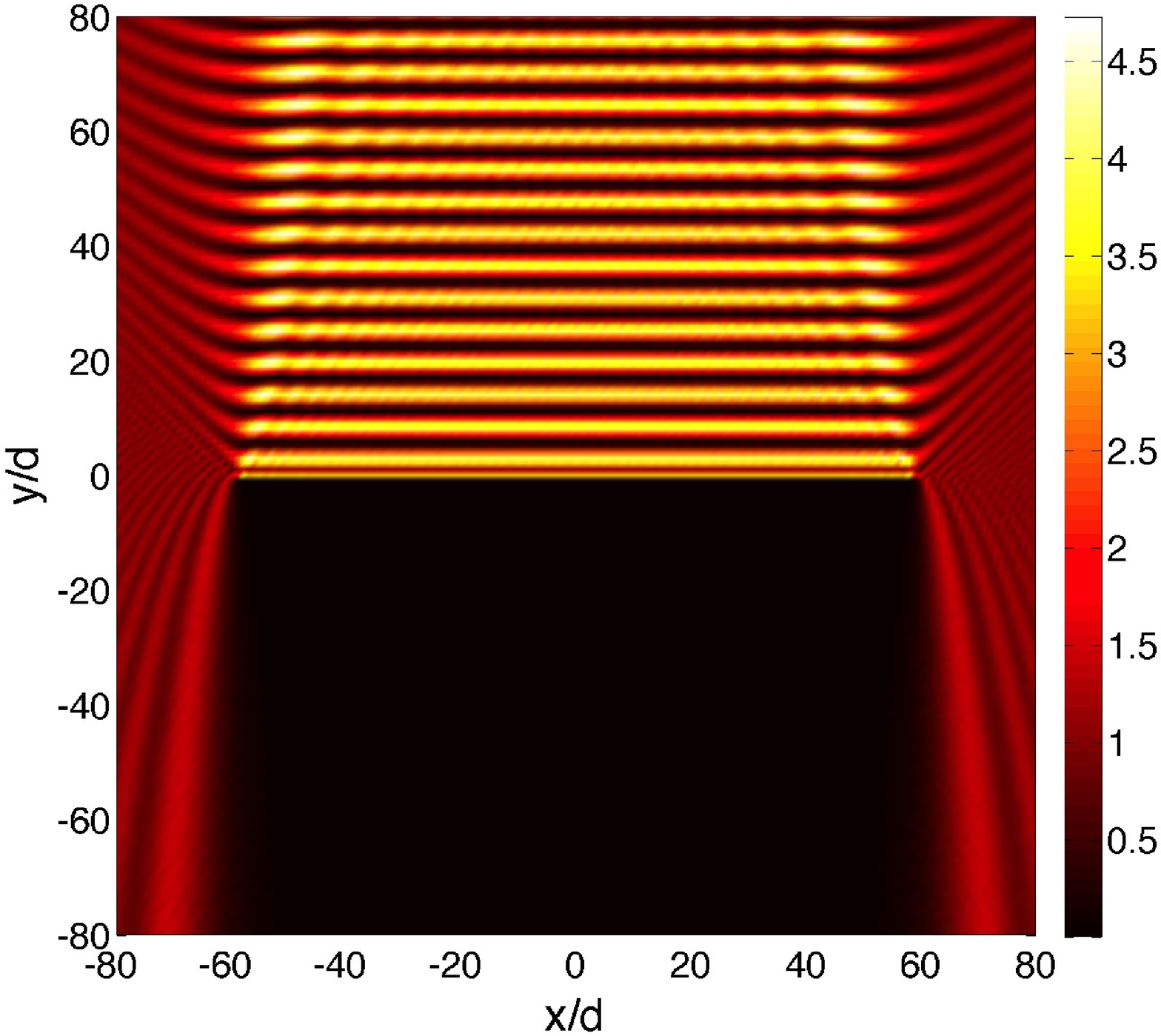}
   \end{center}
   \caption
   { \label{perfect ref} The finite meta-surface is illuminated in normal incidence, from above, by a plane wave at $\omega/\omega_0=0.873$. The map of the square modulus of the electric field is plotted. The map is given in false color with a linear scale.}
\end{figure}
In order to illustrate the Fano behavior, let us plot the map the electric field for $\omega/\omega_0=0.99$ where the transmitted efficiency is equal to $1$, that is the field is perfectly transmitted and $\omega/\omega_0=0.873$, corresponding to a zero of the efficiency, that is a perfectly reflected field. The situation of a perfectly transparent medium is illustrated in fig. \ref{parfaitrans}, where the real part of the electric field is plotted (the energy would only show a uniform color).The map of the energy of the field at the perfect reflection frequency is given in fig.\ref{perfect ref}. The finite meta-surface thus has the property of switching from completely transparent to perfectly reflective with a variation of $10\%$ of the frequency, even with a small number of nano wires.

\begin{figure}
   \begin{center}
   \includegraphics[height=8cm]{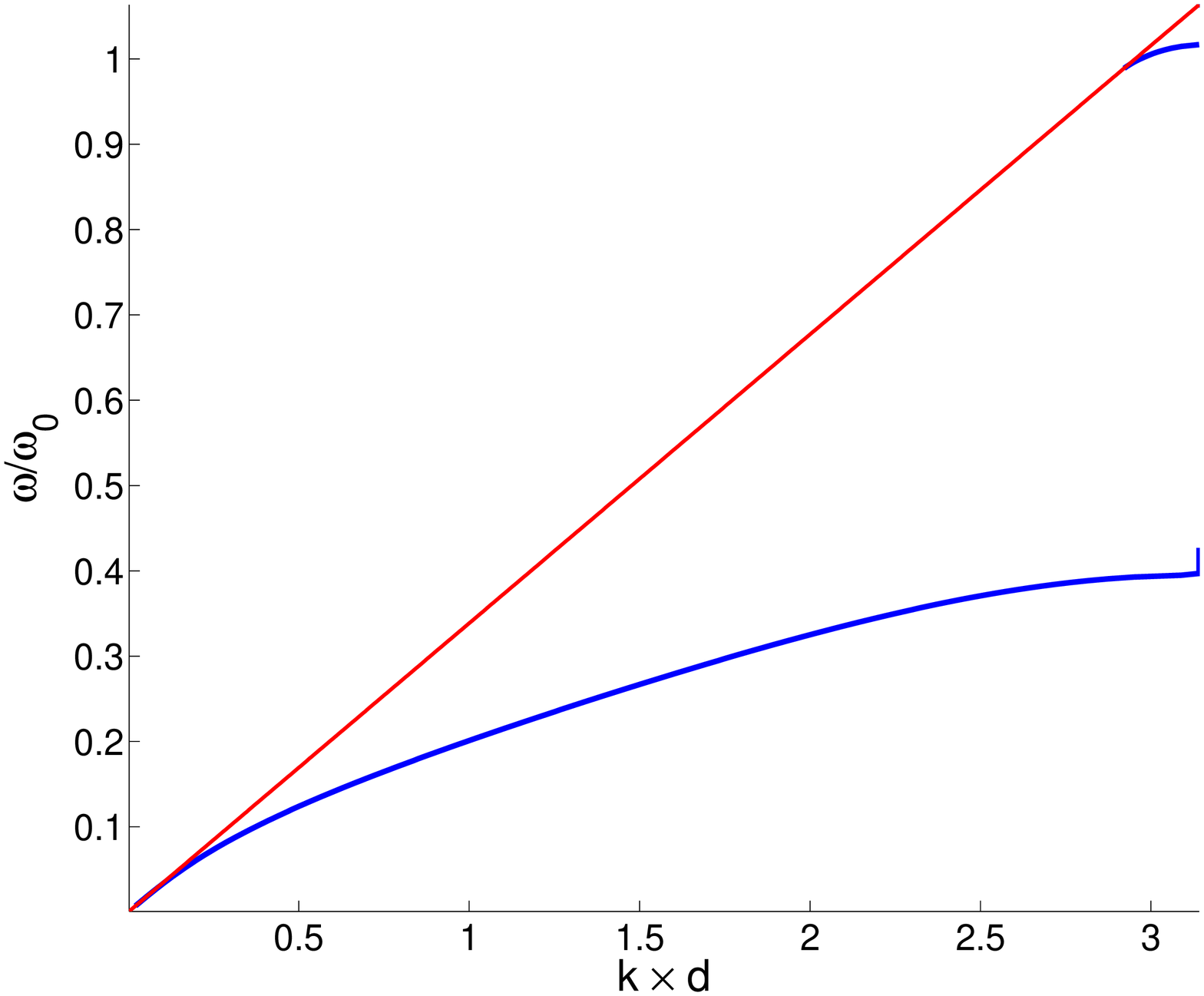}
   \end{center}
   \caption
   { \label{reldisp} Dispersion curves for the meta-surface. The red line corresponds to the light cone. Two bands can be seen: the lower one corresponds to the homogenization regime where the nano wires behaves a homogeneous slab and the upper branch to the Bloch modes corresponding to a collective resonance of a nano wires .}
\end{figure}

The Bloch modes of the meta-surface, obtained from the dispersion relation (\ref{dips}) are given in fig. \ref{reldisp} with respect to the normalized frequency $\omega/\omega_0$. The low-frequency modes correspond to the homogenization regime where the nano wires behave as a dielectric slab.
\begin{figure}
   \begin{center}
   \includegraphics[height=8cm]{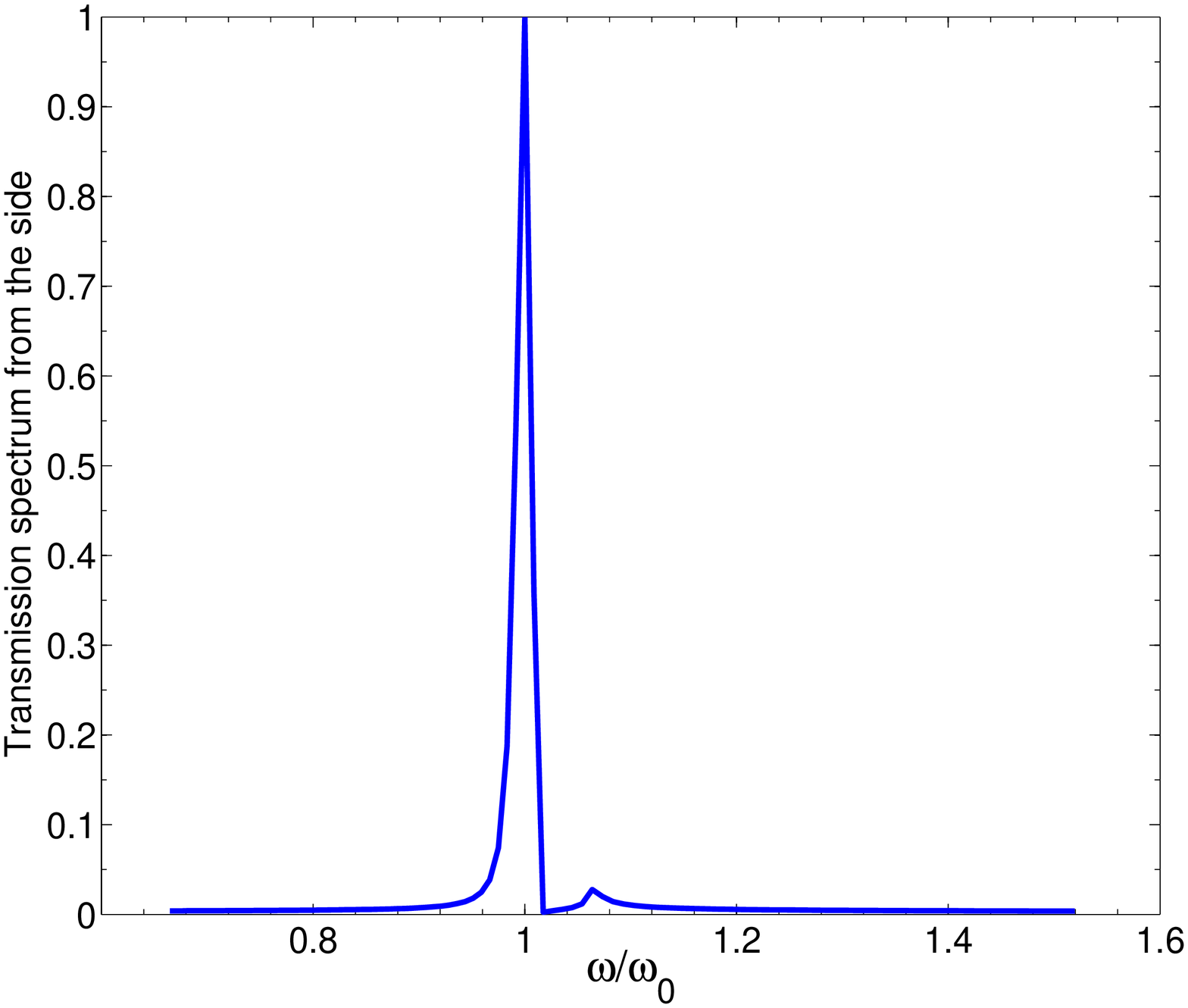}
   \end{center}
   \caption
   { \label{transbloch} Transmission spectrum as a function of the normalized frequency through the meta-surface when illuminated from the side.}
\end{figure}
The band formed above corresponds to the collective behavior of the nano wires near the resonance energy $\omega_0$. These modes are localized in the vicinity of the meta-surface and cannot be excited directly by incident wave.  A possible way to see these modes is to illuminate the nano wires directly from the side (see fig.\ref{transbloch}) and then collect the transmitted field at the other side. The transmission spectrum is given in fig. (\ref{transbloch}).
%Maps of the field corresponding to $\lambda/d=XX$ and $\lambda/d=X$ are given in fig.XX and XX that is for a homogeneous mode and a resonant mode.

%Experimentally, another way of exciting the surface modes is by using the Kreschtmann mount, that is by creating evanescent waves by illuminating a prism in the regime of total internal reflection (see fig. (\ref{transbloch})). This is what is done in fig. XX where we have used $\lambda/d=10$ and $\theta=30∞$.

\begin{figure}
   \begin{center}
   \includegraphics[height=7cm]{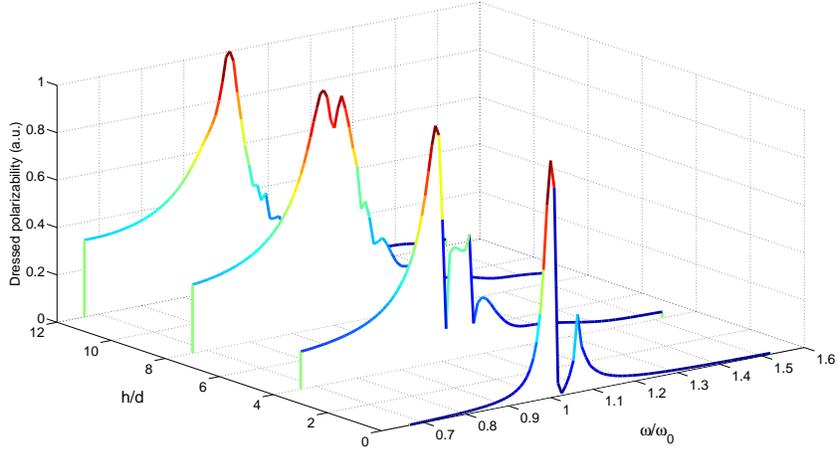}
   \end{center}
   \caption
   { \label{wstrreg} Modulus of the scattering coefficient of a quantum emitter as a function of the frequency and the distance to the meta-surface. The double peaks indicate the strong coupling regime.}
\end{figure}

We turn to the coupling with a quantum emitter. As it has been said before, the emitter is described by a dipole with a frequency-dependent susceptibility. It is a model for a two-level system in the minimal coupling regime. The emitter is situated at some distance $h$ above the meta-surface. The strong and weak coupling regime can be distinguished by evaluating the dressed scattering coefficient of the emitter, that is, the susceptibility modified by the electromagnetic environment.
%When the emitter its spontaneous emission rate is modified by the perturbation of the electromagnetic environnement due to the presence of the meta-surface. This is illustrated in fig. \ref{purcellfac}
%\begin{figure}
%   \begin{center}
%   \includegraphics[height=8cm]{figure6b.eps}
%   \end{center}
%   \caption
%   { \label{purcellfac} Purcell factor as a function of frequency. It represents the modification of the electromagnetic properties of vacuum due to the presence of the meta-surface. It is the ratio between the spontaneous emission rate in the vicinity of the meta-surface over that in vacuum.}
%\end{figure}
This is illustrated in fig. \ref{wstrreg}, where the modulus of the dressed scattering coefficient of the emitter is plotted as a function of both the distance to the meta-surface and the normalized frequency. When the scatterer is far enough from the meta-surface, there is only one peak that exists, corresponding to the Bohr frequency of the emitter. When it gets closer to the meta-surface, a second peak appears, that is due to the coupling with the Bloch mode of the meta-surface, which creates an anti-crossing between the flat dispersion curve of the emitter and that of the Bloch modes supported by the meta-surface.

\section{Conclusion}
A model for the interaction of a dipole with a meta--surface was derived, where the meta-surface is described by an impedance operator. Using this formalism, it is possible to obtain the Bloch electromagnetic surface modes on the meta-surface and to derive their dispersion relation. The Green function and the cross density of states were also derived. Further, the coupling with an emitter characterized by a dressed susceptibility was described and led to the demonstration of the weak and strong coupling regimes with the modes supported with the meta-surface. Because these modes have a very small group velocity, they appear as a peak in the density of states, and, from a quantum point of view, the system can be mapped to a Jaynes-Cummings model. A full quantum model within this approach is currently under study.

\noindent{\bf Acknowledgments}\\
D. Felbacq is an honorary member of the Institut Universitaire de France. Thanks are due to Dr E. Rousseau and to Prof. G. Cassabois for interesting remarks and stimulating discussions.

\end{document}